# Design of Tunable Perfect Absorbers in the Mid-IR Spectrum Using Graphene-Based Multilayer Structures: Emerging Applications in Atmospheric Window Matching

Masoumeh Nazari[1,2], Yaser Mike Banad[1], Sarah S. Sharif, *[1,2]

[1]School of Electrical and Computer Engineering, University of Oklahoma, Norman, OK
[2]Center for Quantum Research and Technology, University of Oklahoma, Norman, OK
*Corresponding Author s.sh@ou.edu

*Abstract*— This paper presents the tunable and switchable Perfect Absorbers (PAs), operating within the mid-infrared (mid-IR) spectrum, specifically targeting the 3 to 5 µm range with precise 0.25 µm intervals. This spectrum is particularly engineered for its minimal atmospheric absorption and unique atmospheric transmission characteristics. Our approach utilizes graphene-based nanophotonic aperiodic multilayer structures optimized through the synergy of micro-genetic optimization algorithm (GOA) within an inverse design framework. This strategic combination enables a predictive model-based strategy that broadens the design of space exploration, facilitating the discovery of PAs with highly accurate absorption control. Employing the Transfer-Matrix-Method (TMM) method for simulations, we manipulate the absorption characteristics, allowing for the precise tailoring of the desired spectral response while maintaining the multilayer structures' thickness under 2 µm. Our results demonstrate the tunability and switchability of PAs by adjusting graphene layers' chemical potentials, highlighting their dynamic optical behavior. For instance, a PA optimized for a 4 µm absorption peak can shift its absorption peak to 4.22 µm by merely changing the graphene layer's chemical potential from 0 eV to 1 eV, without compromising absorption efficiency. Additionally, our research uncovers the proposed absorbers' remarkable adaptability to various incident angles, maintaining 90% absorption up to 52 degrees. This adaptability demonstrates the versatility and robustness of our design across a broad spectrum of real-world applications, including thermal photovoltaics, sensors, and stealth technology, where angular independence significantly enhances device performance and efficiency. This research not only deepens the understanding of nanophotonic materials' capabilities but also paves the way for the design and development of highly efficient optical devices tailored for the mid-IR range.

*Index Terms*—Tunable and switchable perfect absorbers, Mid-infrared spectrum, Graphene-based nanophotonic structure, Inverse design technique.

## I. INTRODUCTION

In recent years, the significance of absorption manipulation has grown exponentially due to the pivotal role of Perfect Absorbers (PAs) in diverse applications. These applications span environmental monitoring [1], thermal emitting [2], sensing [3, 4], photodetection [5], and solar energy harvesting [6], underscoring the versatility and pivotal importance of PAs in advancing technology and addressing global challenges. The utility of PAs have been showcased across diverse frequency ranges in the electromagnetic spectrum, encompassing microwave [7], terahertz (THz) [8], infrared (IR) [9], and visible wavelengths [10]. Notably, the mid-infrared spectrum, which encompasses wavelengths of 3–5 µm, emerges as a focal point of interest. This spectrum is particularly valued for its minimal atmospheric absorption and unique atmospheric transmission characteristics [11], making it an ideal candidate for applications in spectroscopy [12] and sensing [13]. The mid-IR range is especially pertinent for its capacity to detect the absorption lines of specific gases, such as $CO_2$ and methane [14, 15]. This attribute positions PAs as critical tools in the detection and monitoring of greenhouse gases, thereby playing an instrumental role in environmental sustainability initiatives and contributing to the broader efforts in combating climate change.

The spectral attributes of optical absorption, including wavelength, bandwidth, absorption peak, and angular properties, are critically dependent on the careful selection of both materials and design architectures for PAs. The realm of mid-infrared (mid-IR) PAs has seen a proliferation of designs, each distinguished by its unique structural configuration, demonstrating the vast potential for innovation in this field. For instance, Avitzour et al. [16] engineered a complex metamaterial-based PA for the IR region, achieving an impressive absorption efficiency exceeding 90% alongside negligible reflectivity. Similarly, Bossard et al. [17] introduced a polarization-insensitive optical metamaterial absorber with broad bandwidth. Their design, optimized through a genetic algorithm, revealed the complex geometry of a single metallic screen capable of supporting multiple overlapping resonant structures, thereby broadening its applicability. Further advancing the field, Wang et al. [8] showcased a near-perfect absorber tailored for dual-channel operation in the mid-IR, achieving an absorptivity of 99.9% at 25.04 THz. Remarkably, this design was constituted entirely of semiconductor materials, illustrating the potential for semiconductor-based absorbers in high-efficiency applications. Additionally, Chen et al. [18] proposed an ultra-narrowband absorber that integrates a dielectric meta-surface on a metallic film substrate. This design exploits plasmonic absorption within the metal film, with the dielectric meta-surface's super cavity effect significantly enhancing absorption efficiency and achieving band compression.

The versatility of PAs is significantly enhanced by the capacity for dynamic tunability, which enables these devices to flexibly adapt their resonance peaks. Traditional methods for achieving this tunability often rely on the lithographic alteration of absorbers' structural dimensions. While effective, this approach necessitates the complete redesign and fabrication of the absorber for each desired modification, making the process both cumbersome and expensive. In response to these limitations, the scientific community has increasingly focused its efforts on developing tunable PAs, capable of multiband and broadband absorption. These innovative absorbers can dynamically adjust their operational resonance peaks through adjustments in the parameters of the active materials without the need for structural redesign [8, 19-21]. Recent advancements have seen the incorporation of phase-changing materials (PCMs) such as vanadium dioxide ($VO_2$) and germanium antimony telluride (GST) into the design of PAs [22-26]. These materials enable the creation of modulators, switches, and various photonic devices that can be tuned across a wide spectrum of operational states. Nevertheless, a prevalent challenge identified in these configurations is the persistent requirement for high power, limiting their efficacy in low-power photonics integrated devices [26].

In a prior study, we showcased a tunable and switchable thermal emitter using a graphene-based aperiodic multilayer, achieving 100% perfect absorption at 3.34 µm [2]. In this paper, we integrate a novel category of materials into graphene-based aperiodic multilayer structures to propose tunable and switchable PAs, covering an extensive wavelength range, particularly focusing on the mid-IR spectrum. Employing the Transfer-Matrix-Method (TMM) method in conjunction with the micro-genetic algorithm (GA), we meticulously optimize the absorption characteristics at targeted mid-IR wavelengths, simultaneously achieving a significant reduction in layer thickness. Our innovative inverse design strategy allows for the precise customization of PAs, enabling the control of infrared absorption across the entire 3 to 5µm range at fine 0.25 µm intervals. This level of control and flexibility in tuning absorption properties is unparalleled, marking a significant step forward in the development of PAs. Through this approach, we not only enhance the adaptability and efficiency of PAs in various photonic applications but also open new avenues for research in wavelength-specific applications, from environmental sensing to secure communications, thus underscoring the potential of advanced materials and design methodologies in the next generation of optical devices.

Section II of our paper delves into the meticulous selection process of materials integral to the construction of our advanced PAs, accompanied by a detailed explanation of our computational and optimization methodologies. Section III showcases the results obtained from our optimized structure, highlighting its remarkable tunability and switchability characteristics. Additionally, we examine the absorber's performance across various incident angles in both TE and TM polarizations, ensuring its applicability in real-world scenarios. In the concluding section, we summarize key findings and discuss their future implications.

## II. DESIGN AND CHARACTERISTICS ANALYSIS OF GRAPHENE-BASED PERFECT ABSORBER

In this paper, we delve into the design and optimization of aperiodic nanophotonic multilayer structures to achieve peak absorption efficiency within the mid-IR spectrum. Utilizing an advanced inverse design approach, we combine the TMM method with the GOA to optimize the material system. This hybrid approach enables us to precisely calibrate the absorption at specific mid-IR atmospheric windows while minimizing layer thickness to support small size, weight, power, and Cost (SWaP+C) [27]. The structure of the proposed graphene-based mid-IR PA is shown in figure 1, composed of alternating layers of graphene and dielectric, which are sandwiched between Lead Selenide (PbSe) layers.

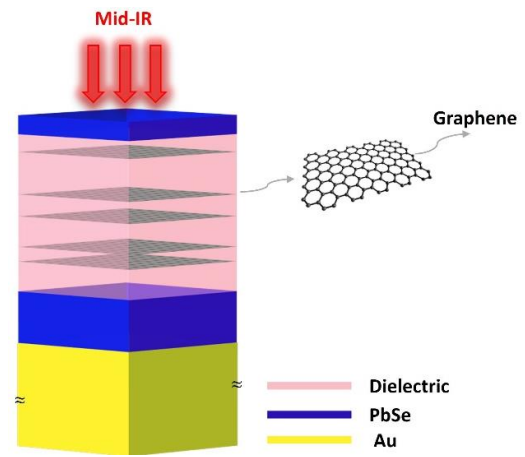

**Fig.1**. The schematic of the proposed Perfect Absorber (PA) composed of alternating layers of graphene and dielectric, which are sandwiched between two PbSe layers with a total thickness of around 2 µm. A semi-infinite gold layer is used as the substrate.

At the core of this advanced multilayer architecture is the incorporation of graphene, a material well-known for its high electron mobility, optical transparency, flexibility, and adjustable conductivity, which efficiently captures incoming mid-IR electromagnetic wave. Graphene, a two-dimensional (2D) sheet comprising carbon atoms arranged in a hexagonal structure, has garnered significant attention owing to its unique electrical, optical, and mechanical properties [28-34]. Furthermore, with its high conductivity in the mid-IR and THz spectrums supporting surface plasmons [35-38], graphene emerges as a promising candidate for diverse applications, encompassing ultrafast photodetectors, modulators, solar cells, and optical absorbers [39-43]. Graphene exhibits efficient absorption primarily owing to its thickness; however, relying solely on the overall absorption of an entire graphene structure may prove inadequate. Therefore, achieving perfect absorption necessitates inventive designs that incorporate graphene-based nanostructures.

An essential component in our design involves Polyphenylsulfone (PPSU), a polymer well-known for its outstanding thermal resilience, serving as an ideal dielectric spacer within the graphene layers [44-46]. PbSe serves as a medium capable of absorbing mid-IR light, and its deposition can be easily achieved through thermal evaporation methods [47]. Nevertheless, despite PbSe's mid-IR light absorption extending up to 4.4 µm [48], achieving perfect absorption would require a significantly thicker structure. Additionally, it's important to note that the absorption of PbSe is not tunable [49, 50]. Our structure incorporates a semi-infinite gold (Au) layer as the substrate, inherently eliminating transmittance and thereby enhancing the absorption efficiency of the PAs. Consequently, $A_{TE/TM}(\lambda) = 1 - R_{TE/TM}(\lambda)$, where $A_{TE/TM}(\lambda)$ is absorptance, $R_{TE/TM}(\lambda)$ is reflectance, and $\lambda$ is the wavelength. Moreover, the reflective properties of gold augment the absorption capability of the graphene layers by creating a semi-mirror structure that enables the light to traverse the layers twice [51]. The interplay between the graphene and dielectric layers supports surface plasmons at their interface, significantly contributing to the absorption mechanism [52]. The graphene and polymer layers can be deposited layer-by-layer to construct a graphene–polymer heterostructure [53], providing accurate control of the spacing between the graphene layers in the proposed aperiodic multilayer structures. This meticulous assembly guarantees the outstanding ability of our structure to manipulate absorption in the 3-5µm range, establishing it as an innovative solution for applications requiring precise control over mid-IR wavelengths. This aperiodic multilayer structure demonstrates spectral-altering properties, acting as a proof of concept for the deployment of more advanced designs optimized for specific applications in the mid-IR spectrum.

Actively managing the propagation of electromagnetic waves is possible through the modification of the chemical potential in graphene. This modification can be accomplished via chemical doping, the application of external electric or magnetic fields, or optical excitation[33]. Therefore, graphene provides a unique opportunity for electrically manipulating the spectral properties of optical absorption. In the proposed structure, the density of charge carriers linked to the chemical potential within the layers of graphene can be controlled by applying a DC bias electric field perpendicular to the surfaces of graphene/dielectric. This results in the electrical manipulation of graphene's refractive index [33]. An appropriate physical parameter for explaining the optical characteristics of graphene is optical conductivity, a complex number linked to the surface current induced in graphene by light [54, 55], which significantly relies on the chemical potential (Fermi energy). The Kubo formula [56] can be employed to model the conductivity of graphene as follows:

$$\sigma(\omega,\mu_c,\Gamma,T) = -\frac{ie^2(\omega+i2\Gamma)}{\pi\hbar^2}\left[\frac{1}{(\omega+i2\Gamma)^2}\int_0^\infty \left(\frac{\partial n_f(\epsilon)}{\partial \epsilon} - \frac{\partial n_f(-\epsilon)}{\partial \epsilon}\right)\epsilon \, d\epsilon - \int_0^\infty \frac{n_f(-\epsilon)-n_f(\epsilon)}{(\omega+i2\Gamma)^2 - 4\left(\frac{\epsilon}{\hbar}\right)^2}d\epsilon\right] \quad (1)$$

where $n_f(\epsilon) = 1/\{1 + \exp[(\epsilon - \mu_c)/(k_B T)]\}$ is Fermi-Dirac distribution, $\omega$ is radian frequency, $e$ is the electron charge, $\hbar$ is reduced Plank constant, T is the temperature, $\mu_c$ is the chemical potential, $k_B$ is the Boltzmann constant, $\Gamma = e\,v_F^2/2\mu_c$ is the charge particle scattering, and $V_F = 10^6$ m/s is the Fermi velocity. The scattering rate for graphene in this context is realistic for multilayer structures, as confirmed by previously conducted relevant experiments [57]. The optical conductivity of graphene is divided into intraband and interband components, corresponding to absorption by free carriers and transitions from the valence band to the conduction band, respectively. In the mid-IR range, the contribution from intraband transitions becomes comparable to that of interband transitions. Consequently, control over intraband transitions, and thus the refractive index, can be achieved by adjusting the chemical potential in graphene [58].

*A. optimization methods*

Our research places a central emphasis on the utilization of inverse design approach that fundamentally differs from traditional design methodologies. Instead of starting with a predefined structure and analyzing its performance, the inverse design approach begins with a desired optical performance goal and works backward to identify the physical structure that achieves this goal, as shown in figure 2(a). This approach is particularly advantageous for our complex multilayer optical systems where the intuitive design may not readily reveal the optimal configurations.

To implement this strategy effectively, we focus on the Genetic Optimization Algorithm (GOA), a type of evolutionary algorithm inspired by the process of natural selection. This algorithm iteratively improves the design by mimicking the processes of mutation, crossover, and selection to explore the vast design space efficiently and identify optimal solutions that meet our criteria for tunable PAs. In our application, the goal is to achieve PAs with specific tunability characteristics within aperiodic multilayer structures. These structures consist of layers with varying thicknesses and materials, presenting a complex optimization problem due to the high dimensionality of the design space and the intricate interactions between layers. To determine the optimal thicknesses for the layers in these aperiodic multilayer structures, we adopted a hybrid optimization approach [59]. This method integrates a micro-genetic algorithm, which is a variant of the genetic algorithm designed for global optimization with a smaller population size for faster convergence, with a local optimization algorithm. The local optimization refines the solutions found by the micro-genetic algorithm, ensuring precise adjustment of the layer thicknesses to achieve the desired absorption characteristics.

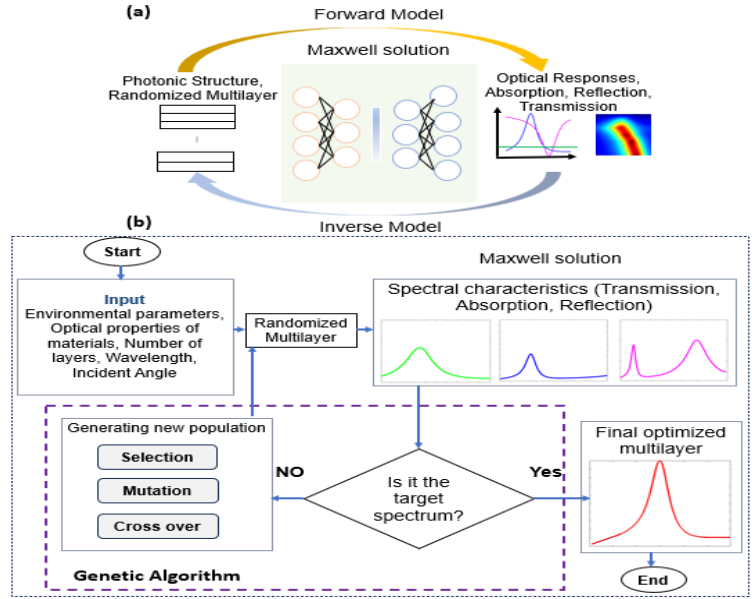

**Fig.2.** The optical response of randomized multilayers is obtained using a forward model, followed by the utilization of an inverse model to generate the optimized multilayers; (b) The optimization approach, integrating the Genetic Optimization Algorithm (GOA) and a local optimization algorithm, strategically refines aperiodic multilayer structures for optimal absorptance. The continuous cycle of stochastic initialization, population evolution, is visually elucidated, highlighting the commitment to maximizing absorptance through the lens of inverse design.

The flowchart depicted in figure 2(b) illustrates the iterative optimization cycle designed to explore the design space. The optimization process begins by generating a population of potential solutions, each distinguished by randomly assigned layer thicknesses. This stochastic initialization serves as the starting point for subsequent iterative optimization. Each iteration refines the solutions further, aiming to pinpoint aperiodic multilayer structures that increase the absorptance of PAs. This iterative optimization process is designed to evolve the population towards superior solutions. The performance of each solution is rigorously evaluated by computing absorptance using the Transfer Matrix Method (TMM) [60], and fitness scores are assigned based on these calculations to quantitatively measure their alignment with the overarching goal of maximizing absorptance. The GOA systematically refines the populations through a sequence of selection, crossover, and mutation operations. This approach, inspired by natural selection, strategically favors solutions with higher fitness scores, introduces genetic diversity through crossover, and facilitates random changes through mutation. The iterative cycle continues until a defined convergence condition is met, resulting in the output of optimal layer thicknesses that decisively maximize the absorptance of the aperiodic multilayer structure at the desired wavelengths.

In parallel with the global GOA, a local optimization algorithm operates, identifying the local optimum within the converged population. This simultaneous fine-tuning process refines the most optimal structure identified through inverse design principles. This innovative hybrid optimization approach, grounded in inverse design principles, underscores its efficacy in determining optimized layer thicknesses for aperiodic multilayer structures and achieving tunable PAs [59, 61, 62].

### III. RESULT AND DISCUSSION

The proposed aperiodic multilayer structures possess the capability to modulate the absorption peak across the entire mid-IR spectrum, achieving absorption peaks at any desired wavelength within the mid-IR range solely through adjustments in layer thickness for the same material systems. Utilizing advanced inverse design methodology, we optimize diverse structures to achieve an absorption peak within the 3-5μm range, making adjustments in 0.25 μm increments. While the materials and number of layers remain consistent for all structures, the optimization algorithm defines the thickness of each layer to maximize absorption in the desired wavelengths, covering the atmospheric windows. Figure 3(a) shows the absorption spectra of nine distinct structures labeled from a to i, each optimized for peak absorption at 3 μm, 3.25 μm, 3.5 μm, 3.75 μm, 4 μm, 4.25 μm, 4.5 μm, 4.75 μm, and 5 μm, respectively. It can be seen that through the interaction of the incident light with the graphene-based nanostructures, all the proposed PAs exhibit almost perfect absorption at desired wavelengths. Figure 3(b) illustrates different aperiodic multilayer absorbers, incorporating alternating layers of dielectric and graphene (represented by the black line), which have been precisely engineered by inverse design to achieve absorption peaks at a particular wavelength. These designs stand out for their compact profiles, maintaining an overall thickness below 2 μm, while each exhibits a unique thickness profile. This variation in thickness plays a pivotal role in enabling the absorbers to selectively target light across different wavelengths. Through adjustments in layer thickness, we can finely tune the absorptive properties, demonstrating the versatility and accuracy of our proposed multilayer absorber design in handling a range of wavelengths.

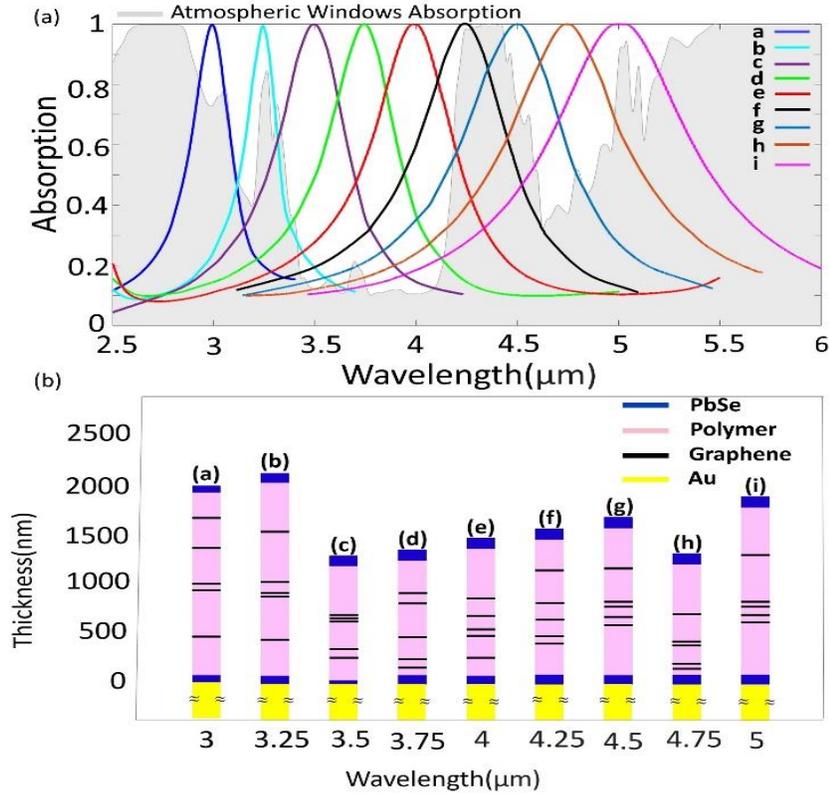

**Fig.3.** (a) Absorption spectra of the proposed absorbers (Structures a, b, c, d, e, f, g, h, and i are optimized to achieve absorption peaks at 3 µm, 3.25 µm, 3.5 µm, 3.75 µm, 4 µm, 4.25 µm, 4.5 µm, 4.75 µm, and 5µm, respectively). (b) Nine optimized aperiodic multilayer absorbers consist of alternating layers of dielectric and graphene each tailored to achieve specific absorption peaks corresponding to (a).

To underscore the significance of the GOA in designing optimal absorbers and elucidate the impact of layer thickness on absorption characteristics, we conduct a comparative analysis between optimized and non-optimized multilayer structures. In figure 4, the absorption spectra vividly depict the remarkable success of the optimized Structure 1, showcasing a meticulous inverse design facilitated by GOA. Conversely, non-optimized Structures 2, 3, and 4, represent varying layer thicknesses derived from the optimized Structure 1, as shown sharp absorption peak precisely at 4 µm, achieved through in the inset figure 4. These structures, featuring non-optimized and arbitrary layer thicknesses of 487.67 nm (maximum), 62.17 nm (minimum), and 274.92 nm (average), distinctly showcase shifts in absorption peaks away from 4 µm and a consequential decrease in absorption efficiency. This comparison underscores the significance of employing GOA for the engineering of multilayer structures, emphasizing its transformative impact on absorption characteristics for specific wavelength tuning.

To illustrate the dynamic tunability and switchability in our design, we investigated the impact of varying chemical potentials on the graphene layers within our structures figure 5 depicts the absorption spectra of nine optimized structures as a function of the chemical potentials. For structures (a) and (b), an elevation in the chemical potential from 0 eV to 1 eV results in a shift of the absorption peak from 3 µm to 3.05 µm and from 3.25 µm to 3.29 µm, respectively. Consequently, the interplay of graphene-dielectric layers and PbSe layers' absorption contributes to tunability, showcasing outstanding and nearly flawless absorption characteristics in these specific wavelengths The average contribution of graphene to absorption in these structures hovers around 20%, with PbSe predominantly responsible for the majority of absorption. When comparing these results with those of the other optimized structures depicted in figure 5 (c), (d), (e), (f), (g), and (h), it becomes evident that the impact of varying chemical potential on absorption peaks is more notable at longer wavelengths. This highlights the substantial tunability and switchability of the proposed structures, suggesting a significant contribution from the graphene layers to the absorption rate. For instance, figure 5(e), optimized specifically for peak absorption at 4 µm, illustrates the structure's dynamic tunability: at a chemical potential of 0 eV, it achieves perfect absorptance (unity) at 4 µm. With an increase in the chemical potential to 1 eV, the peak absorptance shifts to 4.22 µm while maintaining a high absorptance of 0.9. This highlights the structure's remarkable capability to sustain high absorption efficiency while demonstrating a noticeable tunability range.

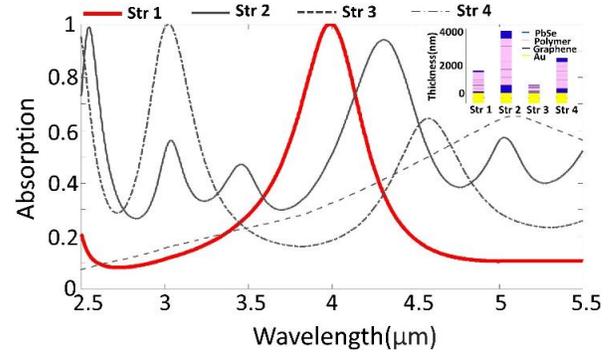

**Fig.4.** Impact of the Genetic Optimization Algorithm (GOA) on the absorption characteristics of multilayer structures. Absorption spectra for optimized and non-optimized structures, highlighting the sharp absorption peak at 4 µm achieved by the optimized Structure 1 through precise inverse design, and non-optimized Structures 2, 3, and 4, each representing varying layer thicknesses of 487.67 nm (maximum), 62.17 nm (minimum), and 274.92 nm (average) based on the optimized Structure 1. The inset figure illustrates the layer thickness distribution for each structure, underscoring the tailored thickness in the optimized structure versus the varied and non-ideal thicknesses in the non-optimized ones.

The switchable characteristic of our structure is exemplified by modulating perfect absorption through variations in chemical potential at the optimized wavelength. Specifically, in figure 5(e), optimized for a peak absorption at 4 µm, substantial switchability is observed: transitioning from a chemical potential of 0 eV to 1 eV results in a decrease in absorption at 4 µm from 100% to 56%, emphasizing its notable switchable behavior. The optimized structure presented in figure 5(i) further illustrates this switchability, demonstrating a significant shift. Remarkably, the perfect absorptance, initially achieving unity at 5 µm with a chemical potential of zero, can be adjusted to an absorptance of 0.55 by changing the chemical potential to 1 eV. This underscores the practicality of designing Perfect Absorbers (PAs) that are both tunable and switchable. Previous research has shown that switchability is enhanced by adding more graphene layers. Although our designs incorporate a relatively modest number of graphene layers, they still manifest significant switchability, demonstrating the effectiveness of our approach in achieving dynamic control over absorption properties.

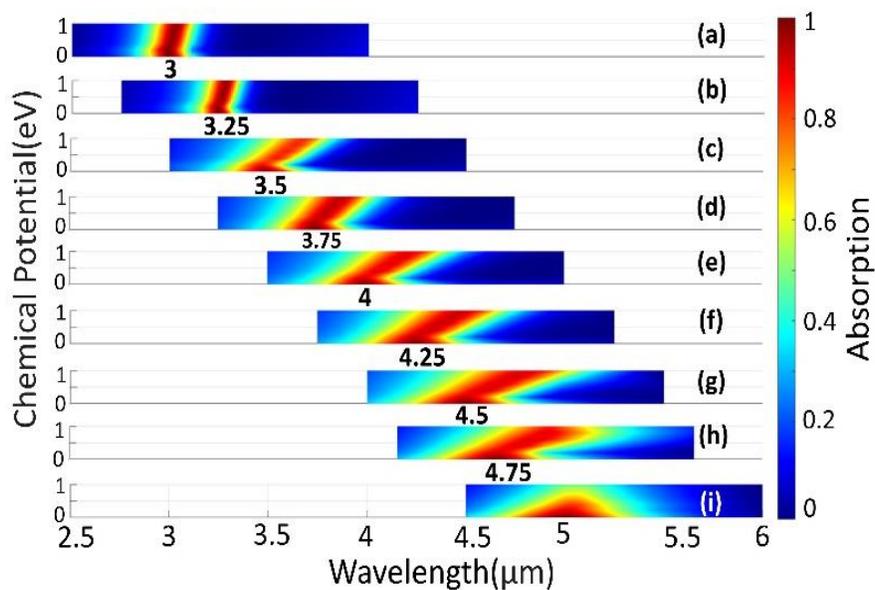

**Fig.5**. The impact of varying chemical potential on the absorption characteristics of the proposed Perfect Absorbers (PAs). (a) and (b) A light redshift occurs with an increase in chemical potential, accompanied by achieving perfect absorption at these specific wavelengths; (c), (d), (e), (f), (g), (h) Tunable absorption through chemical potential modulation. Increasing the chemical potential leads to a redshift in absorption peaks towards longer wavelengths. (i) Switchable absorption through chemical potential modulation. The perfect absorptance of unity at 5 µm for a chemical potential of 0 can be switched to an absorptance of 0.55 by adjusting the chemical potential to 1 eV.

Figure 6(a) depicts the normalized profile of the electric field amplitude concerning the incident plane wave's amplitude at 4 µm wavelength, showcasing adjustments in chemical potential within the optimized structure designed for the absorption peak at 4 µm. Notably, upon setting the chemical potential at $\mu_c = 0.0$ eV for optimal absorptance, the electric field amplitude for normal light incidence maintains a near-flat profile in the air. This characteristic signifies an almost negligible reflectance in the structure, thereby leading to an absorptance approaching unity.

In figure 6(b), the influence of absorbing materials (Graphene and PbSe) in the proposed structure is presented for $\mu_c = 0.0$ eV and 1.0 eV, emphasizing the switchability inherent in the proposed structures. Notably, the contribution of graphene layers to absorption in the proposed structure experiences a significant reduction with an increase in chemical potential. At $\mu_c = 0.0$ eV, graphene layers contribute 48% to absorption. However, at $\mu_c = 1.0$ eV, this contribution diminishes to around 2%. As the energy absorbed in the multilayer structure is proportionate to the refractive index [63], this connection can be influenced by adjusting the chemical potential of graphene. The noticed modifications in optical absorption characteristics, resulting from changes in the chemical potential of graphene layers, confirm the possibility of creating optical absorbers that can be controlled electrically.

In our study, although we have effectively optimized our structures for perfect absorption under normal incident light, we recognize the significance of preserving high absorption efficiency when exposed to different incident angles for practical applications. In response to this concern, we have conducted a comprehensive examination of how the incident angle impacts the performance of our proposed PAs. Figure 7 presents a comprehensive overview of absorption spectra at various incident angles, spanning from θ=0° (normal incidence) to θ=90° (grazing incidence), for all optimized multilayer structures designed to absorb within the 3 to 5 µm range, with 0.25 µm intervals. This analysis reveals an interesting characteristic: as the incident angle increases, there is a noticeable shift in the absorption peaks towards shorter wavelengths. Despite these shifts, the proposed absorbers demonstrate remarkable resilience in maintaining high absorption levels. The absorption rate exceeds 90%, even at incident angles as steep as approximately 46, 47, 49, 50, 52, 55, 57, 58, and 60 degrees for structures (a), (b), (c), (d), (e), (f), (g), (h), and (i) respectively. These findings hold importance as they validate the robustness and applicability of our PA designs. The capacity to maintain nearly perfect absorption across a broad range of incident angles not only underscores the effectiveness of our structures but also underscores their suitability for practical applications, where light commonly approaches from diverse angles.

To illustrate the inherent controllability and tunability of our design, we analyze the effect of varying chemical potential and incident angle on a specific structure (e). highlights how changes in chemical potential and incident angle influence the absorptive capacity of this structure, optimized for peak absorption at 4 µm. This figure provides a clear visualization of the structure's dynamic adjustability, showing that altering the chemical potential from 0 eV to 1 eV shifts the absorption peak at λ=4 µm towards higher wavelengths under normal incidence while varying the incident angles for constant chemical potentials causes the absorption peak to shift towards lower wavelengths. Nevertheless, it is evident that alterations of both chemical potential and incident angle lead to a decline in absorption performance. Despite this, the structure maintains a high absorption peak above 0.9 for angle adjustments up to 52 degrees. This demonstrates the proposed structure's capability to function as a near-perfect absorber (PA) in practical situations where light incidence angles can vary.

Figure 9 illustrates the absorption spectrum for both TE and TM polarizations, showcasing the impact of varying the incident angle for structure (e). It is evident that the absorption performance is superior for TM polarization compared to TE polarization. The structure can operate close to a PA with 90% absorption, allowing adjustments in the angle up to 60 degrees for TM polarization, while reaching a limit of 47 degrees for TE polarization. This difference underscores the structure's ability to handle a wider range of incident angles effectively when the magnetic component of the light is more engaged (TM polarization), providing greater flexibility in applications where the angle of incidence cannot be precisely controlled.

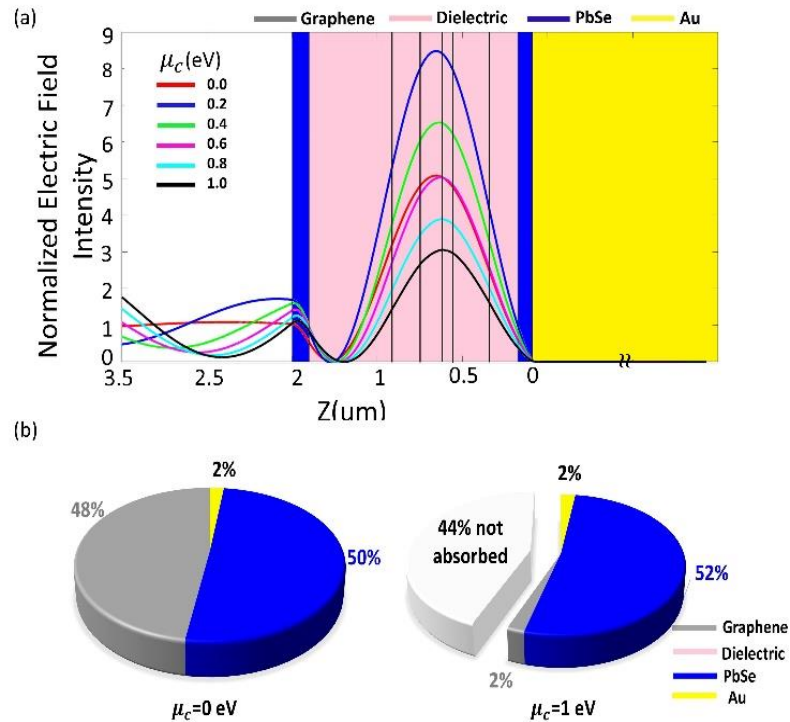

**Fig.6**. (a) Normalized electric field amplitude profile for Structure (e), optimized for the absorption peak at 4 µm. At a chemical potential of $\mu_c$ = 0.0 eV, wherein the structure is fine-tuned for maximum absorptance, the electric field amplitude remains nearly constant in air, exhibiting an almost flat characteristic. This outcome implies that the structure has almost no reflectance, leading to an absorptance that is close to unity. (b) Contribution of Graphene and PbSe to overall absorption in multilayer Structure (e). At $\mu_c$ = 0.0 eV, graphene layers absorb 48% of the incident light, contrasting with a reduced 2% when varying $\mu_c$ = 1.0 eV, illustrating the structure's switchability.

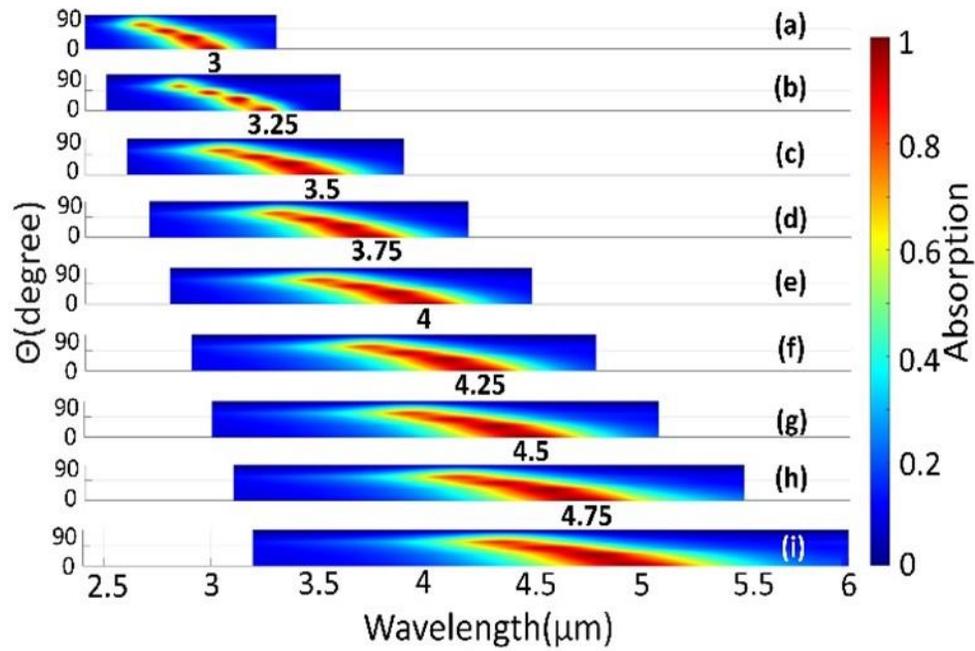

**Fig.7**. The effect of incident angle on absorption of the proposed PAs. The observation reveals a blue shift in absorption peaks as incident angles increase across all structures. These proposed structures consistently sustain their PA status, achieving over 90% absorption, up to incidence angles of 46, 47, 49, 50, 52, 55, 57, 58, and 60 degrees for structures (a), (b), (c), (d), (e), (f), (g), (h), and (i), respectively.

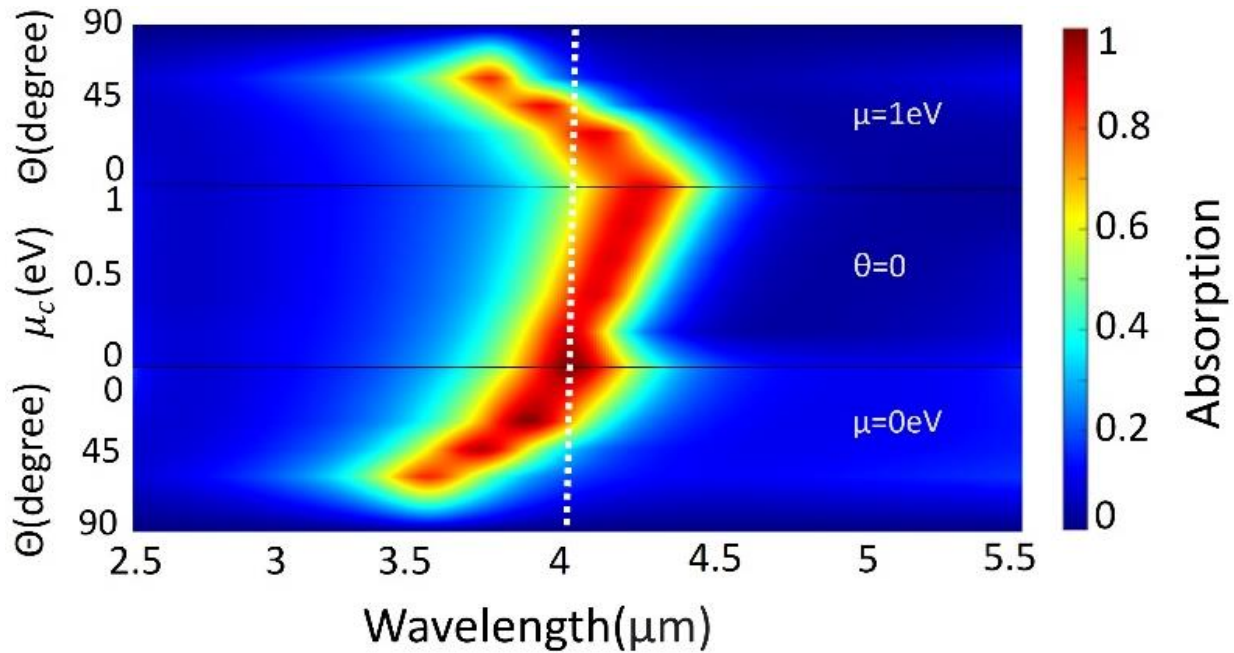

**Fig.8.** The impact of altering the chemical potential and incident angle on the absorption spectrum of structure (e), optimized for an absorption peak at 4 μm. The absorption peak of this structure can be tuned from λ=4 μm to λ=4.22 μm by varying the chemical potential from 0 eV to 1 eV. Remarkably, there is substantial switchability, resulting in a decrease in absorption at 4 μm from 100% to 56% during the transition from 0 eV to 1 eV. The Structure still has more than 90% absorption by adjusting the angle up to 52°. changing in chemical potential and incident angle at the same time results in a deterioration of absorption performance.

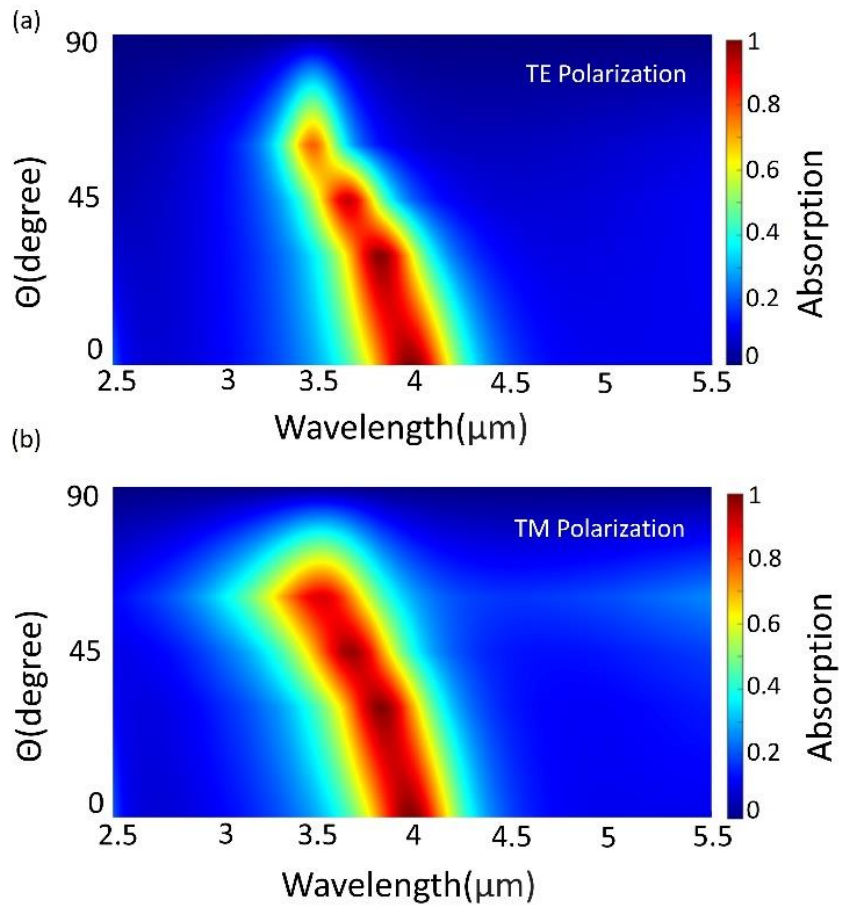

**Fig.9.** The impact of varying the incident angle on the absorption spectrum for structure (e), optimized for absorption peak at 4 μm, for both (a) TE; and (b) TM polarizations. The structure exhibits superior absorption performance in TM polarization compared to TE polarization. Specifically, it functions as a PA with the flexibility to adjust the angle up to 60 degrees in TM polarization, while the limit is 47 degrees for TE polarizations.

## Conclusion

This study introduces a novel methodology for crafting tunable and switchable Perfect Absorbers (PAs) within the mid-IR spectrum (3-5 μm). Employing graphene-based nanophotonic aperiodic multilayer structures optimized through a micro-genetic optimization algorithm (GOA) in an inverse design framework, our approach effectively addresses the challenges associated with designing PAs for multilayer structures targeting specific wavelengths. The application of GOA is pivotal, enabling precise engineering of layer thicknesses, facilitating the absorption of any desired wavelength, and navigating optimization complexities within multilayer optical systems. Importantly, by meticulously controlling the atmospheric window within the 3-5 μm range through the design of structures that efficiently absorb light in this specific spectral band, our study adds a critical dimension to its significance, holding profound implications for applications reliant on mid-infrared technology. This capability enhances the adaptability and effectiveness of the proposed PAs in real-world scenarios. The distinctive feature of our proposed designs lies in their exceptional tunability and switchability, achieved by manipulating the chemical potential of graphene through the application of various bias voltages. This adaptability in manipulating wavelengths and absorptance signifies a substantial advancement in mid-IR absorber technology. Furthermore, the proposed PAs exhibit remarkable adaptability to varying incident angles, maintaining high absorption efficiency up to 60 degrees. the research expounded in this paper not only introduces a pioneering tunable absorber design but also provides a framework for future advancements in mid-IR photonic applications. Striking a balance between material efficiency and functional versatility, this work contributes significantly to the field, opening avenues for innovative applications in mid-infrared technology.